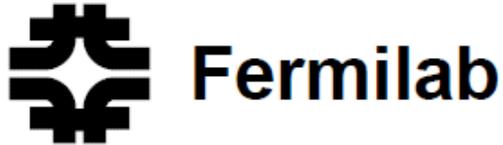

# Activation assessment of the soil around the ESS accelerator tunnel[*]

**I. L. Rakhno**[1][†]**, N. V. Mokhov**[1]**, I. S. Tropin**[1]**, D. Ene**[2]

[1]Fermi National Accelerator Laboratory, Batavia, Illinois 60510, USA
[2]European Spallation Source, ESS-ERIC, Tunavagen 24, SE-221 00 Lund, Sweden

## Abstract

Activation of the soil surrounding the ESS accelerator tunnel calculated by the MARS15 code is presented. A detailed composition of the soil, that comprises about 30 different chemical elements, is considered. Spatial distributions of the produced activity are provided in both transverse and longitudinal direction. A realistic irradiation profile for the entire planned lifetime of the facility is used. The nuclear transmutation and decay of the produced radionuclides is calculated with the DeTra code which is a built-in tool for the MARS15 code. Radionuclide production by low-energy neutrons is calculated using the ENDF/B-VII evaluated nuclear data library. In order to estimate quality of this activation assessment, a comparison between calculated and measured activation of various foils in a similar radiation environment is presented.

# Activation assessment of the soil around the ESS accelerator tunnel

I L Rakhno[1], N V Mokhov[1], I S Tropin[1], D Ene[2]

[1] Fermi National Accelerator Laboratory, Batavia, Illinois 60510, USA
[2] European Spallation Source, ESS-ERIC, Tunavagen 24, SE-221 00 Lund, Sweden

E-mail: rakhno@fnal.gov

**Abstract**. Activation of the soil surrounding the ESS accelerator tunnel calculated by the MARS15 code is presented. A detailed composition of the soil, that comprises about 30 chemical elements, is considered. Spatial distributions of the produced activity are provided in both transverse and longitudinal directions. A realistic irradiation profile for the entire planned lifetime of the facility is used. The nuclear transmutation and decay of the produced radionuclides is calculated with the DeTra code which is a built-in tool for the MARS15 code. Radionuclide production by low-energy neutrons is calculated using the ENDF/B-VII evaluated nuclear data library. In order to estimate quality of this activation assessment, a comparison between calculated and measured activation of various foils in a similar radiation environment is presented.

## 1. Introduction

The ESS linear accelerator is planned to be operating at an average beam power of 5 MW for 40 years [1]. Various radionuclides will be produced in the surrounding soil and surface water by secondary particles (mostly neutrons) generated in the machine components and target by the proton beam with energy up to 2 GeV. Therefore, a detailed analysis is needed to predict residual activity of the soil and water outside the accelerator tunnel and target station. This study addresses distributions of activity of radionuclides produced in the soil around the ESS accelerator tunnel wall for two periods of time: one year and forty years of normal operation for a given operational scenario. Further spatial migration of the produced radionuclides in the soil is not considered in this paper and will be a subject of another dedicated study.

Only the contribution from the beam losses in the linac (from beginning to the end of the accelerator to the target region, up to a neutron shield wall) is considered here. The primary particles are not tracked beyond the rightmost shielding wall of the accelerator tunnel.

The detailed distributions of the produced radionuclides are provided in 30 spatial bins—3 longitudinal and 10 transverse bins—for a further migration analysis by a dedicated expert group. Total residual activity of the accelerator tunnel itself is calculated as well.

A brief description of computer codes used for the calculations is provided. In order to verify quality of the predicted residual activation, one compares calculated and measured residual activity of various foils irradiated in a similar radiation environment. The comparison reveals a good agreement between the measurements and calculations.

## 2. MARS15 and DeTra computer codes

*2.1. MARS15 code*

MARS15 is a general-purpose Monte Carlo code for modeling particle and heavy ion interactions with matter and charged particle transport in magnetic field in realistic accelerator three-dimensional structures [2, 3, 4]. The energy span covers many decades: from a multi-TeV region down to the thermal energies for neutrons and to keV region for other particles. The code has more than a 40-year history and for the last approximately 25 years is developed and maintained at Fermi National Accelerator Laboratory. In our energy region of interest—below 2 GeV—the following data libraries and nuclear collision models are used:

- Below 100 MeV (adjustable), TENDL-2015 library for protons, light ions and gammas, and CEM model (code) for all other projectiles;
- Mix-and-match procedure around 100 MeV between TENDL and CEM model (code);
- Exclusive modeling with CEM code for energies between 100 and 300 MeV;
- Mix-and-match procedure between CEM and LAQGSM model (code) between 300 and 500 MeV;
- Exclusive modeling with LAQGSM above 500 MeV.

Neutron interactions below 14 MeV are modeled using ENDF/B-VII library.

At present, DeTra code is implemented as a part of MARS15 code, and it can be invoked separately after MARS itself generates output files with calculated nuclide production rates in specified regions of interest.

*2.2. DeTra code*

DeTra code has approximately a 20-year history and was developed at Helsinki University and CERN for various calculations of residual activity at accelerators [5]. As the name suggests, the code performs decay, build-up and transmutation calculations by means of analytically solving Bateman equations for a given irradiation and cooling down scenario. The complexity of the decay chains and the number of nuclides are not limited as long as the nuclear data are known (supplied). At present, library NUDAT 2.6 is used as a source of nuclear decay data.

**3. Geometry model of the ESS accelerator and tunnel with service buildings**

The MARS15 computation model developed and described in detail in Refs. [6, 7] is used for normal operational beam losses for calculations presented in this paper. Only the major beam line components were modelled, and support structures and devices were excluded from the calculation model. The right-hand coordinate system origin (x=y=z=0) is at the beginning of the ESS linac, with x-axis pointing up, y-axis - to the right and z-axis along the beam in the linac horizontal section. In some cases, the s-axis coinciding with the beam axis is used. Figure 1 shows a vertical scan at y=0 through the ESS accelerator MARS15 model while a plan view at x=0 (beam level) is given in Fig. 2. A soil berm above the accelerator tunnel has a noticeable right (positive y) to left (negative y) slope with the highest vertical coordinate being x=9.3 m. The berm height is at x=8.42m above the beam (y=0), with the overall ground level at the site at x=2.1 m.

The radionuclide production is scored in transverse bins in soil immediately outside the concrete tunnel structures. The first five bins are 20-cm thick each, while the bins 6 thru 10 are 100-cm thick each. In addition, in order to provide more detailed data, the linac length was divided into three regions:

1. The low-beta region, with z from 0 to 176 m, with the corresponding maximum proton energy of 571 MeV;
2. the region downstream the region 1 and up to the end of the linac straight section, with z from 176 to 490.56 m, with the corresponding proton energies from 571 MeV to 1.991 GeV;
3. the region downstream the region 2 and up to the end of the accelerator, with z from 490.56 up to 580.5 m, with the proton energy of 1.991 GeV.

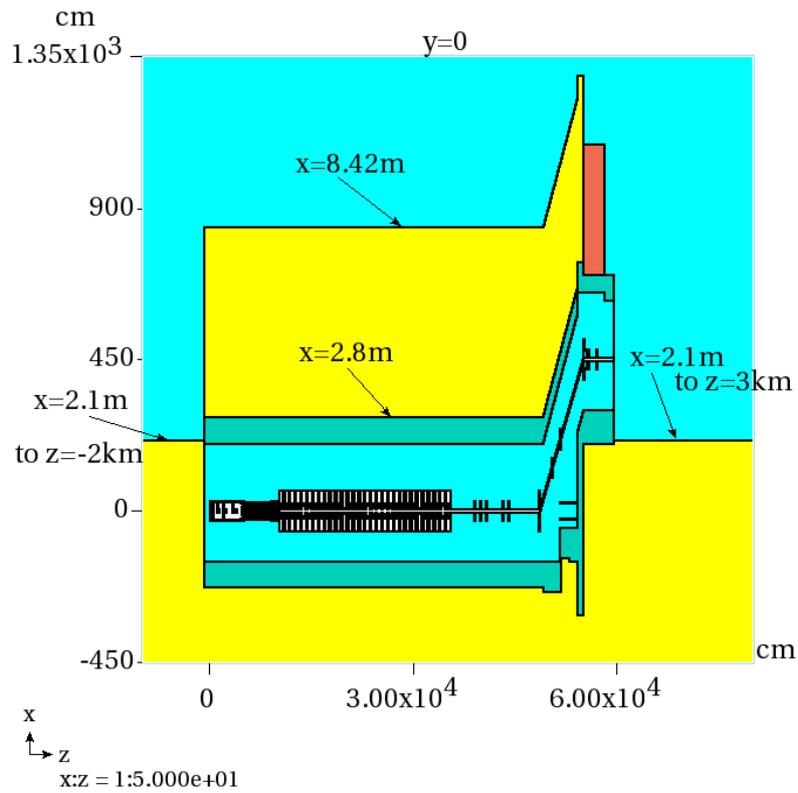

**Figure 1.** A side view of the ESS accelerator MARS15 model [6, 7].

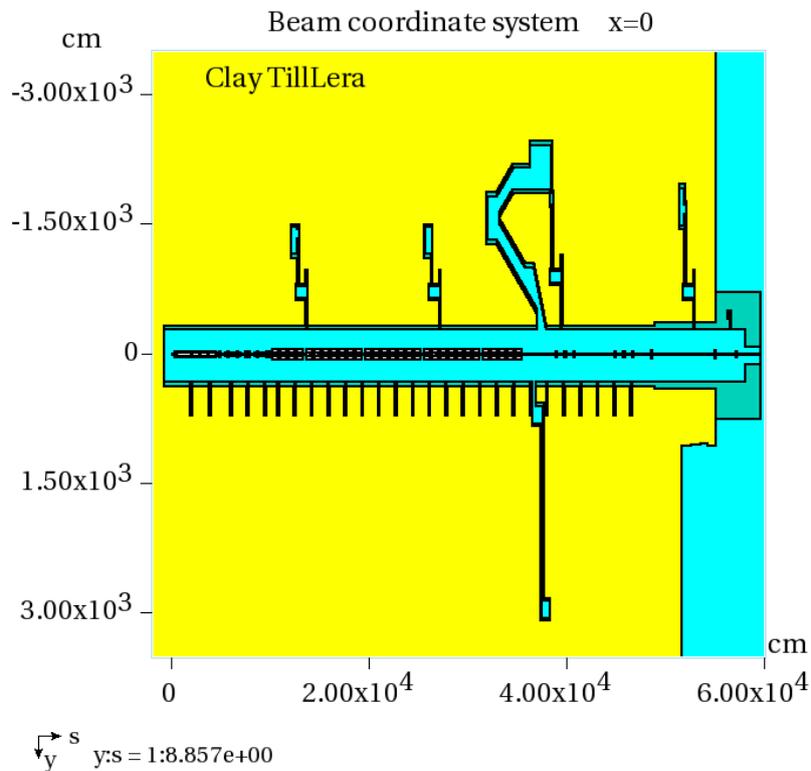

**Figure 2.** A plan view of the ESS accelerator and adjacent structures in MARS15 calculation model [6, 7]. Here the label "s" refers to the beam line coordinate.

Several fragments of the MARS15 model geometry, that are specific to this study on produced radionuclides, are given in Figs. 3-7. The various colours shown in the bins are used for technical reasons to specify volumes for scoring radionuclide production in the corresponding materials.

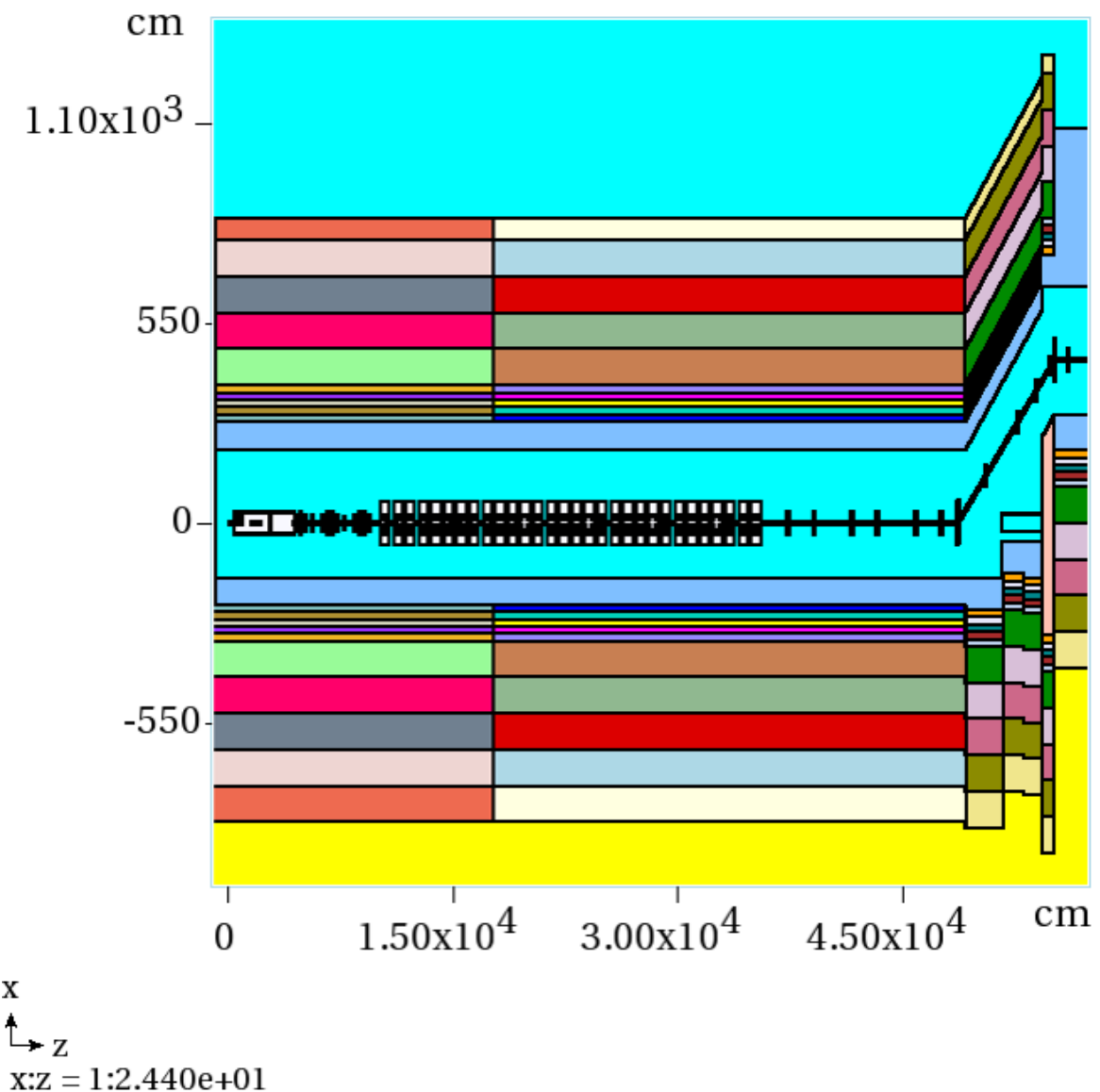

**Figure 3.** A side view of the ESS accelerator and adjacent structures in the MARS15 calculation model with transverse bins for scoring produced radionuclides.

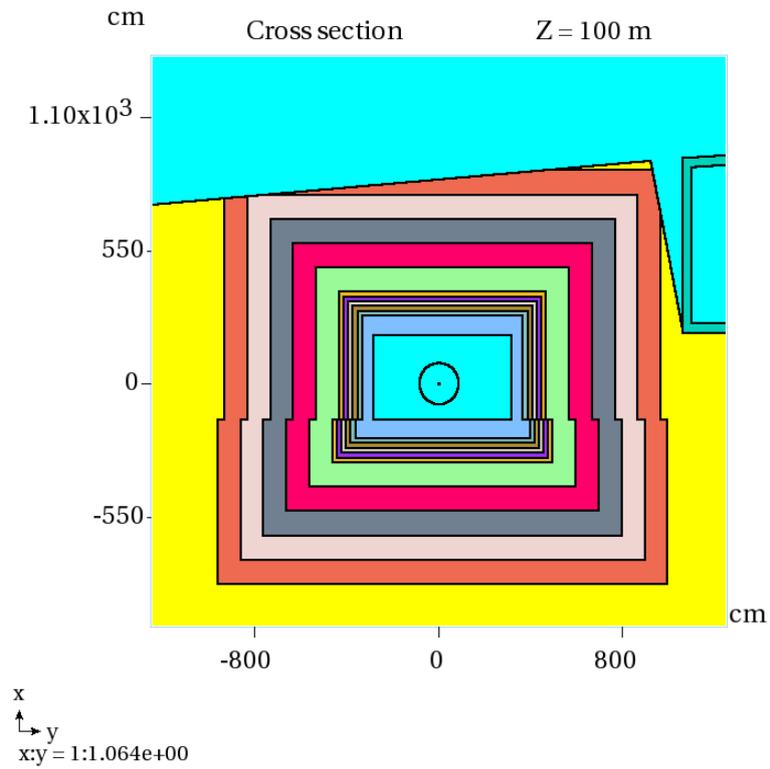

**Figure 4.** A cross section of the ESS accelerator and adjacent structures in the MARS15 calculation model in region 1 at z=100m with transverse bins for scoring produced radionuclides.

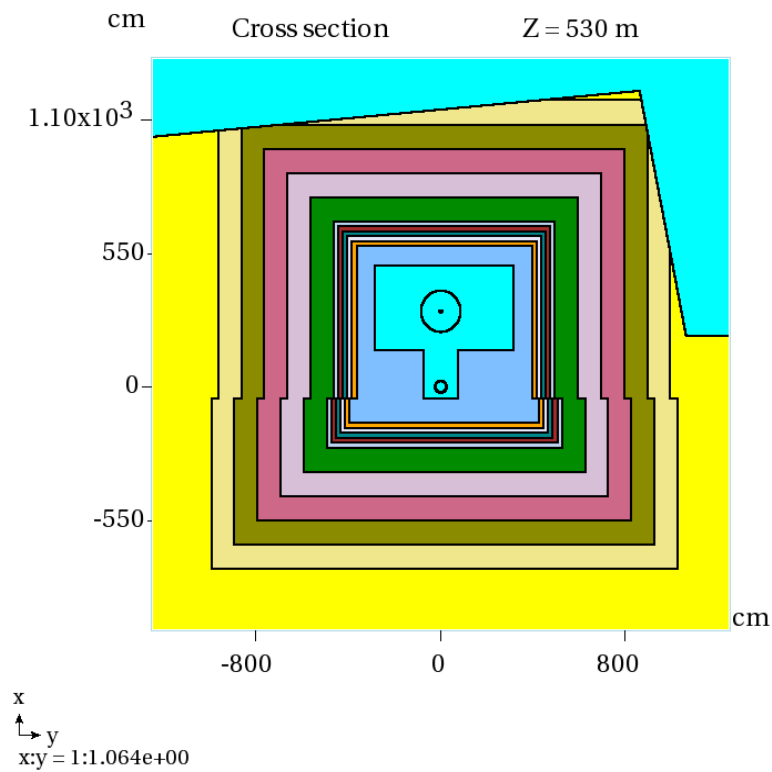

**Figure 5.** A cross section of the ESS accelerator and adjacent structures in the MARS15 calculation model in region 3 at z=530m with transverse bins for scoring produced radionuclides.

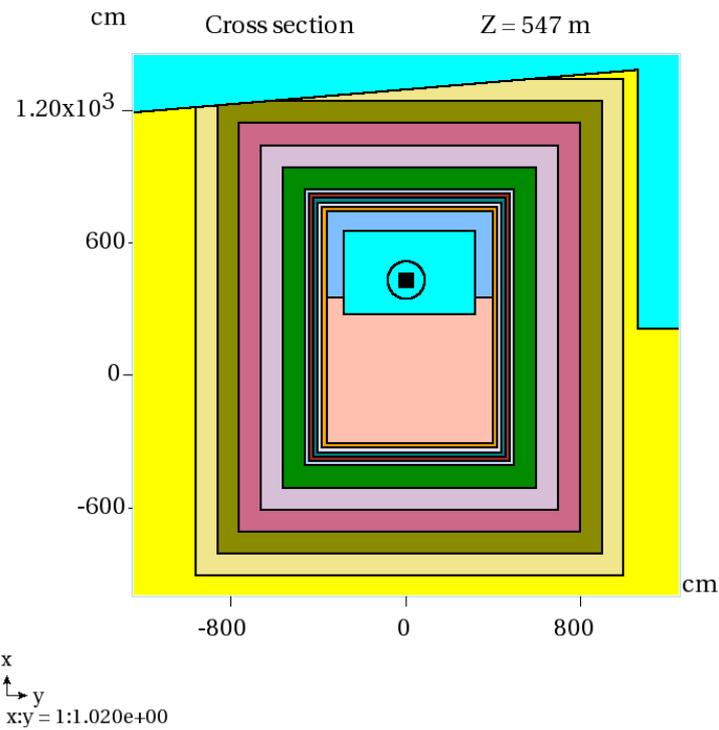

**Figure 6.** A cross section of the ESS accelerator and adjacent structures in the MARS15 calculation model in region 3 at *z*=547m with transverse bins for scoring produced radionuclides (the pink region underneath the blue region is the same concrete wall which, for technical reasons, serves as a placeholder for the beam dump in our model, see also Fig. 3).

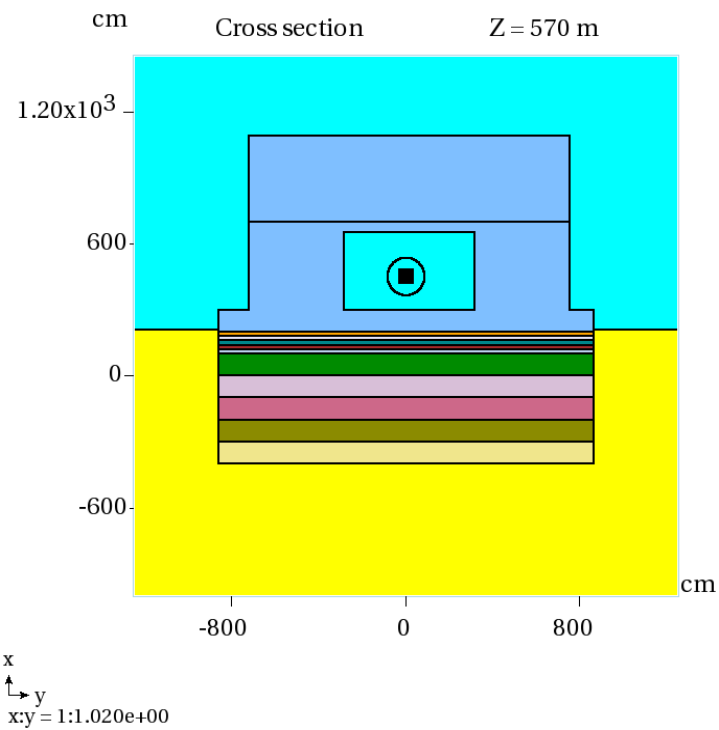

**Figure 7.** A cross section of the ESS accelerator and adjacent structures in the MARS15 calculation model in region 3 at *z*=570m with transverse bins for scoring produced radionuclides (the dark blue regions represent concrete).

## 4. Assumptions and input data

### 4.1. Operational scenario
The following operational scenario is assumed: in every single year, a 6000-hour beam-on period is followed by a 2760-hour beam-off period. For the 1st and 40th year of operation, the nuclide production is calculated immediately after the most recent beam-on period (in other words – at shut-down).

### 4.2. Beam loss
At present, for high-energy proton accelerators it is a common practice to use the so-called "1 W/m" rule derived from the hands-on maintenance conditions and suggested in Ref. [8]. It provides an average operational beam loss along the entire linac beam line except for the very first low-energy sections. The 1 W/m average homogeneous beam loss rate has been adopted at ESS [9] and is considered as a design criterion and an upper limit during normal operations.

The beam energy dependence on location along the ESS accelerator and the modelled beam loss rate at 1 W/m are shown in Fig. 8.

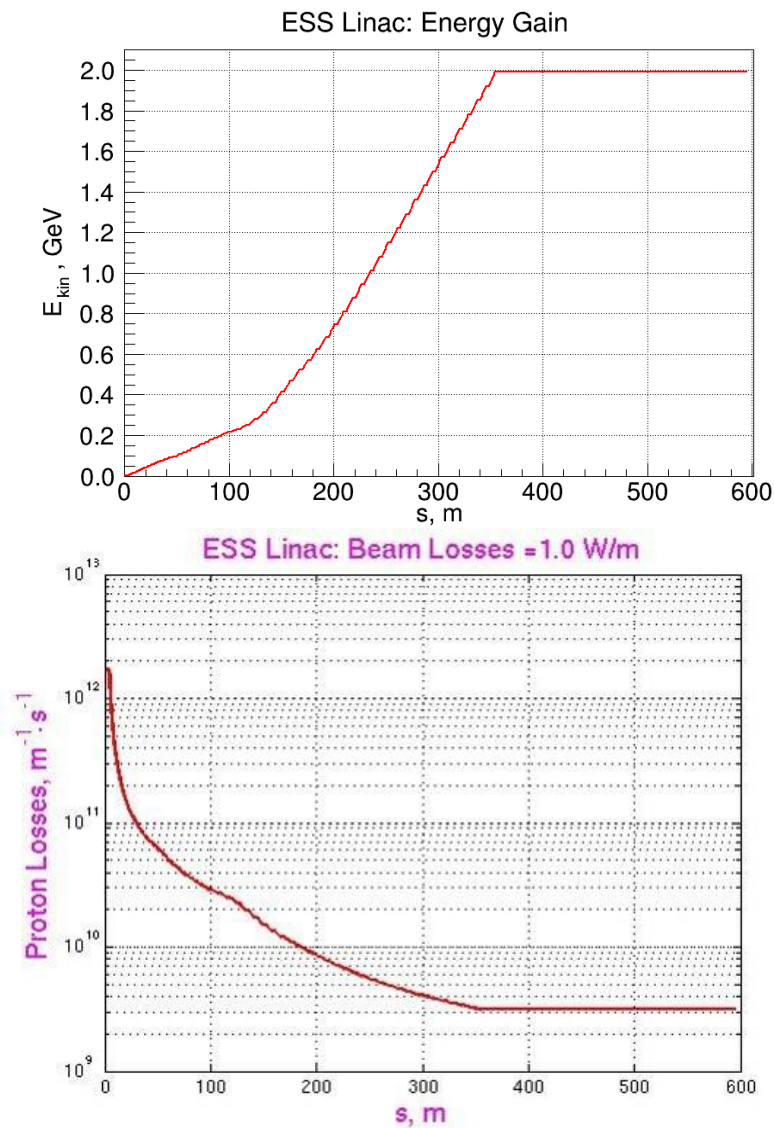

**Figure 8.** Primary proton beam energy as a function of location along the ESS linac (top) and the modelled normal operational beam loss rate at 1 W/m [6, 7] (bottom).

*4.3. Shielding materials*

Ordinary concrete and soil are the two main shielding materials of ESS accelerator. Chemical composition and density of both used at ESS are given in Table 1. The composition of the soil used in calculations is assumed to be a 50%-50% mixture of the deep and shallow soil sample compositions. All other materials in the MARS15 model have their standard compositions and densities. These include air, stainless steel, niobium, liquid helium, titanium, aluminium, magnet yoke steel, cables, copper, water and stub filler (90% sand mixed with 10% of poly-beads).

**Table 1.** ESS accelerator shielding material properties. Density of both deep and shallow ESS soil is equal to 2.0 g/cm$^3$, and density of ESS concrete is equal to 2.35 g/cm$^3$.

| Chemical component | ESS Shallow Soil Sample Weight fraction (%) | ESS Deep Soil Sample Weight fraction (%) | ESS Concrete Weight fraction (%) |
|---|---|---|---|
| Water | 16.4901 | 16.0827 | 3.00 |
| Mg | 0.5156 | 0.8278 | 0.19 |
| Al | 4.4515 | 3.6631 | 1.92 |
| Si | 15.45 | 15.45 | 12.51 |
| S | 0.00 | 0.00 | 0.81 |
| K | 2.0085 | 2.0389 | 1.58 |
| Ca | 0.8828 | 6.1268 | 39.27 |
| Ti | 0.2149 | 0.1740 | 0.38 |
| Mn | 0.0317 | 0.0283 | 0.21 |
| Fe | 2.2578 | 1.9147 | 4.44 |
| Rb | 0.0107 | 0.0095 | 0.00 |
| Ba | 0.0327 | 0.0274 | 0.00 |
| Ni | 0.0017 | 0.0015 | 0.12 |
| Zr | 0.0068 | 0.0055 | 0.00 |
| O | 57.00 | 53.00 | 35.57 |
| H | 1.8452 | 1.7997 | 0.00 |
| Ce | 0.0052 | 0.0044 | 0.00 |
| Co | 0.0009 | 0.0008 | 0.00 |
| Cr | 0.0040 | 0.0033 | 0.00 |
| Cu | 0.0004 | 0.0005 | 0.00 |
| Dy | 0.0005 | 0.0003 | 0.00 |
| Eu | 0.0001 | 0.0001 | 0.00 |
| La | 0.0025 | 0.0020 | 0.00 |
| Na | 0.5630 | 0.5706 | 0.00 |
| Nd | 0.0022 | 0.0023 | 0.00 |
| P | 0.0482 | 0.0399 | 0.00 |
| Pb | 0.0015 | 0.0008 | 0.00 |
| Sc | 0.0007 | 0.0095 | 0.00 |
| Sr | 0.0054 | 0.0158 | 0.00 |
| V | 0.0050 | 0.0041 | 0.00 |
| Y | 0.0016 | 0.0013 | 0.00 |
| Yb | 0.0002 | 0.0001 | 0.00 |
| Zn | 0.0039 | 0.0029 | 0.00 |

## 5. Results of calculations

The list of eleven most important radionuclides for groundwater activation was compiled using studies from Ref. [10]. This list comprises the following nuclides: $^{3}$H, $^{7}$Be, $^{22}$Na, $^{24}$Na, $^{32}$P, $^{35}$S, $^{45}$Ca, $^{46}$Sc, $^{54}$Mn, $^{55}$Fe, $^{65}$Zn. In this paper, the calculated activation is provided for the radionuclides from this list. At the same time, output files of these calculations contain information on all produced radionuclides.

In these calculations, statistical RMS uncertainty, 1σ, of the production rate for $^{3}$H, $^{7}$Be, $^{22}$Na and $^{24}$Na does not exceed 1%. $^{65}$Zn represents produced nuclides with the lowest activity, and corresponding 1σ is approximately 8%. For the remaining produced nuclides on the list, the values of 1σ are in between 1 and 8%.

Table 2 provides the total longitudinally-integrated activity (Bq) of the relevant accounted eleven nuclides for the first transverse 1-m thick soil layer and the one for the entire soil, the 6-m thick layer. Results are given for 1 year and 40 years of operation. Table 3 provides results for the specific activity (Bq/kg) for the same scoring scheme. Table 4 provides the total activity (Bq) in the 6-m thick soil layer for the first, second and third longitudinal regions as well as the sum of the three regions after one year of operation.

The total activity over all the lateral bins in the soil, as well as total activity of the tunnel concrete and dump placeholder concrete are shown in Figure 9 and Table 5. The top ten radionuclides in the total soil volume after 40 years of operation at shut down and after 30 years of cooling are given in Table 6. The spatial distribution in the soil of the specific activity follows that of the total hadron flux above 30 MeV. Figs 10 through 12 shows lateral distributions of such a flux in the three longitudinal sections of the linac. Energy spectra of neutrons and gammas, averaged over all the ten transverse bins of the soil and over the tunnel concrete in the 3$^{rd}$ longitudinal section, are shown in Fig. 13.

**Table 2.** Activity (Bq) of radionuclides produced in the layers of soil – 1-m thick and 6-m thick – around the ESS tunnel wall immediately after 1 year and 40 years of operation (no decay).

| Produced nuclide | Period of operation (yr) | | | |
|---|---|---|---|---|
| | 1 | | 40 | |
| | Layer thickness (m) | | Layer thickness (m) | |
| | 1 | 6 | 1 | 6 |
| $^{3}$H | 1.157×10$^{9}$ | 1.285×10$^{9}$ | 1.892×10$^{10}$ | 2.101×10$^{10}$ |
| $^{7}$Be | 6.631×10$^{9}$ | 7.340×10$^{9}$ | 6.688×10$^{9}$ | 7.404×10$^{9}$ |
| $^{22}$Na | 9.188×10$^{8}$ | 1.022×10$^{9}$ | 3.930×10$^{9}$ | 4.371×10$^{9}$ |
| $^{24}$Na | 4.635×10$^{10}$ | 5.008×10$^{10}$ | 4.635×10$^{10}$ | 5.008×10$^{10}$ |
| $^{32}$P | 1.469×10$^{9}$ | 1.608×10$^{9}$ | 1.470×10$^{9}$ | 1.609×10$^{9}$ |
| $^{35}$S | 4.482×10$^{8}$ | 4.976×10$^{8}$ | 4.746×10$^{8}$ | 5.268×10$^{8}$ |
| $^{45}$Ca | 3.053×10$^{7}$ | 3.396×10$^{7}$ | 3.870×10$^{7}$ | 4.305×10$^{7}$ |
| $^{46}$Sc | 8.595×10$^{8}$ | 9.276×10$^{8}$ | 9.036×10$^{8}$ | 9.752×10$^{\backslash 8}$ |
| $^{54}$Mn | 8.088×10$^{8}$ | 8.998×10$^{8}$ | 1.457×10$^{9}$ | 1.620×10$^{9}$ |
| $^{55}$Fe | 5.883×10$^{8}$ | 6.531×10$^{8}$ | 2.630×10$^{9}$ | 2.919×10$^{9}$ |
| $^{65}$Zn | 1.093×10$^{6}$ | 1.209×10$^{6}$ | 1.693×10$^{6}$ | 1.873×10$^{6}$ |

**Table 3.** Specific activity (Bq/kg) of radionuclides produced in the layers of soil – 1-m thick and 6-m thick – around the ESS tunnel wall immediately after 1 year and 40 years of operation (no decay).

| Produced nuclide | Period of operation (yr) | | | |
|---|---|---|---|---|
| | 1 | | 40 | |
| | Layer thickness (m) | | Layer thickness (m) | |
| | 1 | 6 | 1 | 6 |
| $^3$H | 34.4 | 3.95 | 563 | 64.6 |
| $^7$Be | 197 | 22.6 | 199 | 22.8 |
| $^{22}$Na | 27.3 | 3.14 | 117 | 13.4 |
| $^{24}$Na | 1380 | 154 | 1380 | 154 |
| $^{32}$P | 43.7 | 4.95 | 43.8 | 4.95 |
| $^{35}$S | 13.3 | 1.53 | 14.1 | 1.62 |
| $^{45}$Ca | 0.91 | 0.104 | 1.15 | 0.132 |
| $^{46}$Sc | 25.6 | 2.85 | 26.9 | 3.0 |
| $^{54}$Mn | 24.1 | 2.77 | 43.4 | 4.98 |
| $^{55}$Fe | 17.5 | 2.01 | 78.3 | 8.98 |
| $^{65}$Zn | 0.033 | 0.0037 | 0.05 | 0.0058 |

**Table 4.** Longitudinal distribution of total activity (Bq) of radionuclides produced in the 6-m thick soil layer around the ESS tunnel wall immediately after 1 year of operation (no decay).

| Produced nuclide | 0<Z<176m | 176m<Z<490.56m | 490.56m<Z<580.5m | Total |
|---|---|---|---|---|
| $^3$H | $5.147 \times 10^7$ | $1.097 \times 10^9$ | $1.367 \times 10^8$ | $1.285 \times 10^9$ |
| $^7$Be | $2.065 \times 10^8$ | $6.340 \times 10^9$ | $7.936 \times 10^8$ | $7.340 \times 10^9$ |
| $^{22}$Na | $3.957 \times 10^7$ | $8.740 \times 10^8$ | $1.083 \times 10^8$ | $1.022 \times 10^9$ |
| $^{24}$Na | $4.395 \times 10^9$ | $4.105 \times 10^{10}$ | $4.633 \times 10^9$ | $5.008 \times 10^{10}$ |
| $^{32}$P | $1.041 \times 10^8$ | $1.345 \times 10^9$ | $1.591 \times 10^8$ | $1.608 \times 10^9$ |
| $^{35}$S | $2.357 \times 10^7$ | $4.215 \times 10^8$ | $5.252 \times 10^7$ | $4.976 \times 10^8$ |
| $^{45}$Ca | $1.325 \times 10^6$ | $2.934 \times 10^7$ | $3.297 \times 10^6$ | $3.396 \times 10^7$ |
| $^{46}$Sc | $8.464 \times 10^7$ | $7.575 \times 10^8$ | $8.544 \times 10^7$ | $9.276 \times 10^{\backslash 8}$ |
| $^{54}$Mn | $4.269 \times 10^7$ | $7.631 \times 10^8$ | $9.397 \times 10^7$ | $8.998 \times 10^8$ |
| $^{55}$Fe | $3.242 \times 10^7$ | $5.521 \times 10^8$ | $6.853 \times 10^7$ | $6.531 \times 10^8$ |
| $^{65}$Zn | 0 | $1.069 \times 10^6$ | $1.401 \times 10^5$ | $1.209 \times 10^6$ |

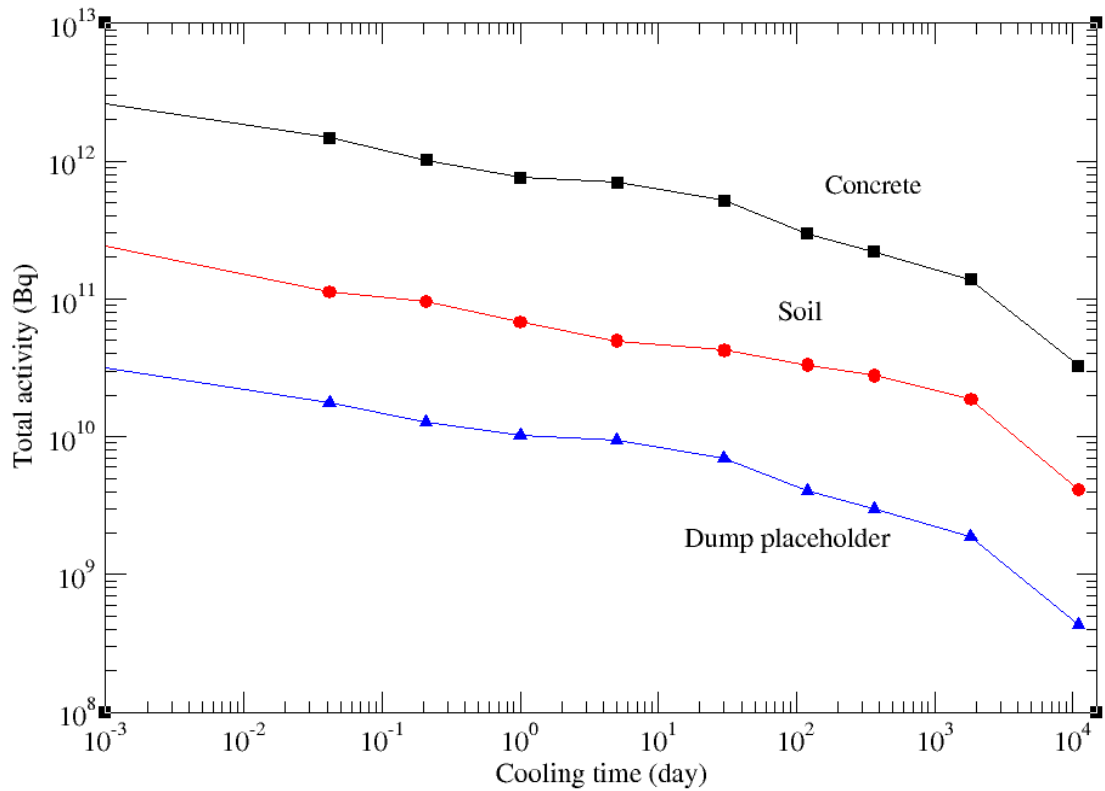

**Figure 9.** Total activity of all the bins in the soil, as well as total activity of the tunnel concrete and dump placeholder concrete (see Figs. 3 and 6) after 40 years of operation. Cooling time 0 corresponds to the moment of the accelerator shut down.

**Table 5.** Total activity (Bq) of all the bins in the soil, as well as total activity of the tunnel concrete and dump placeholder concrete after 40 years of operation.

| Cooling time | Tunnel concrete | Total soil | Dump placeholder concrete |
| --- | --- | --- | --- |
| 0 | $3.744 \times 10^{12}$ | $4.019 \times 10^{11}$ | $4.649 \times 10^{10}$ |
| 1 hour | $1.484 \times 10^{12}$ | $1.126 \times 10^{11}$ | $1.751 \times 10^{10}$ |
| 5 hours | $1.015 \times 10^{12}$ | $9.597 \times 10^{10}$ | $1.286 \times 10^{10}$ |
| 1 day | $7.583 \times 10^{11}$ | $6.821 \times 10^{10}$ | $1.026 \times 10^{10}$ |
| 5 days | $7.007 \times 10^{11}$ | $4.958 \times 10^{10}$ | $9.487 \times 10^{9}$ |
| 30 days | $5.182 \times 10^{11}$ | $4.251 \times 10^{10}$ | $7.005 \times 10^{9}$ |
| 120 days | $2.978 \times 10^{11}$ | $3.322 \times 10^{10}$ | $4.014 \times 10^{9}$ |
| 1 year | $2.204 \times 10^{11}$ | $2.789 \times 10^{10}$ | $2.967 \times 10^{9}$ |
| 5 years | $1.382 \times 10^{11}$ | $1.879 \times 10^{10}$ | $1.866 \times 10^{9}$ |
| 30 years | $3.265 \times 10^{10}$ | $4.133 \times 10^{9}$ | $4.344 \times 10^{8}$ |

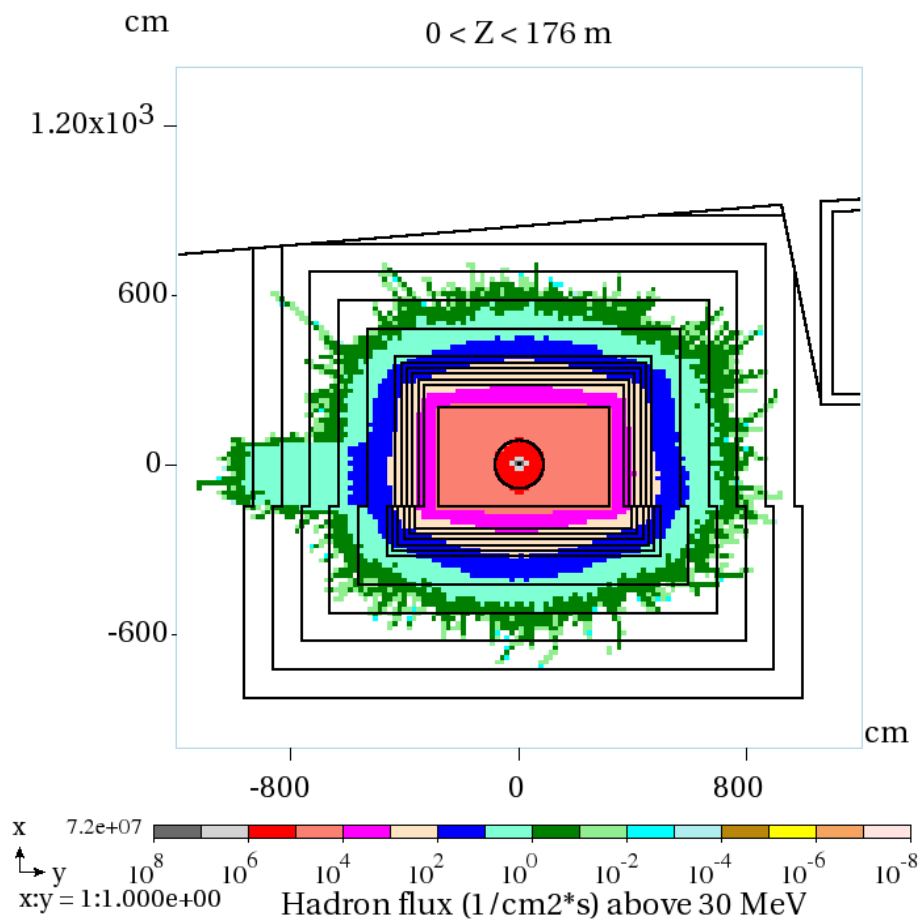

**Figure 10.** Transverse distribution of total hadron flux above 30 MeV averaged over the first longitudinal section of the model.

**Table 6.** Activity (Bq) of top ten radionuclides in all the soil bins after 40 years of operation given at shut down (left) and after subsequent 30 years of cooling (right).

| Produced radionuclide | Activity | Produced radionuclide | Activity |
| --- | --- | --- | --- |
| $^{15}$O | $1.076 \times 10^{11}$ | $^{3}$H | $3.994 \times 10^{9}$ |
| $^{28}$Al | $1.035 \times 10^{11}$ | $^{39}$Ar | $4.722 \times 10^{7}$ |
| $^{24}$Na | $5.140 \times 10^{10}$ | $^{14}$C | $4.159 \times 10^{7}$ |
| $^{3}$H | $2.157 \times 10^{10}$ | $^{41}$Ca | $1.907 \times 10^{7}$ |
| $^{27}$Si | $1.356 \times 10^{10}$ | $^{60}$Co | $1.102 \times 10^{7}$ |
| $^{13}$N | $1.213 \times 10^{10}$ | $^{44}$Sc | $4.059 \times 10^{6}$ |
| $^{56}$Mn | $1.120 \times 10^{10}$ | $^{44}$Ti | $4.059 \times 10^{6}$ |
| $^{14}$O | $8.259 \times 10^{9}$ | $^{22}$Na | $1.525 \times 10^{6}$ |
| $^{7}$Be | $7.599 \times 10^{9}$ | $^{55}$Fe | $1.511 \times 10^{6}$ |
| $^{11}$C | $7.168 \times 10^{9}$ | $^{133}$Ba | $1.425 \times 10^{6}$ |

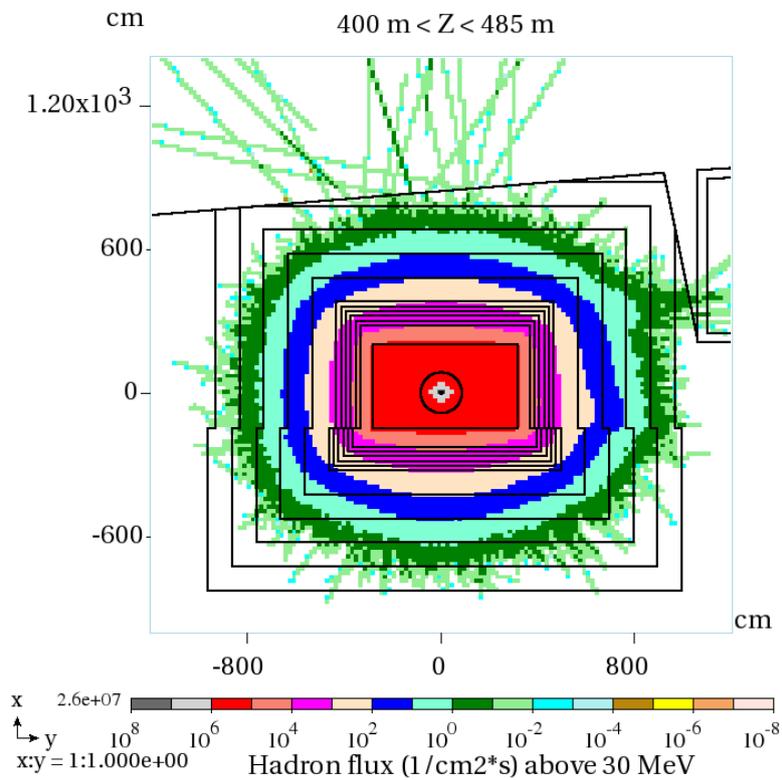

**Figure 11.** Transverse distribution of total hadron flux above 30 MeV averaged over a part the second longitudinal section of the model.

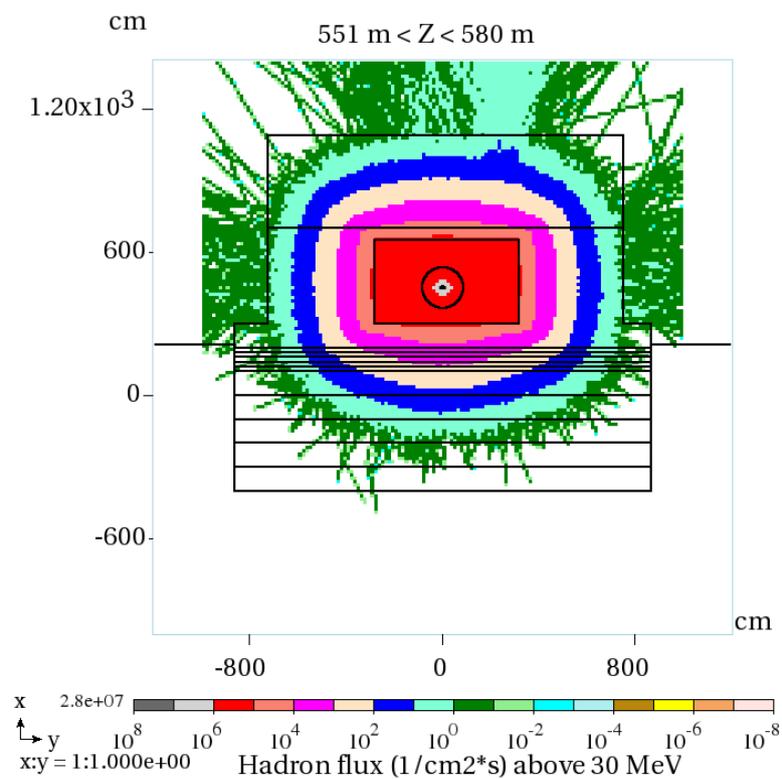

**Figure 12.** Transverse distribution of total hadron flux above 30 MeV averaged over a part the third longitudinal section of the model.

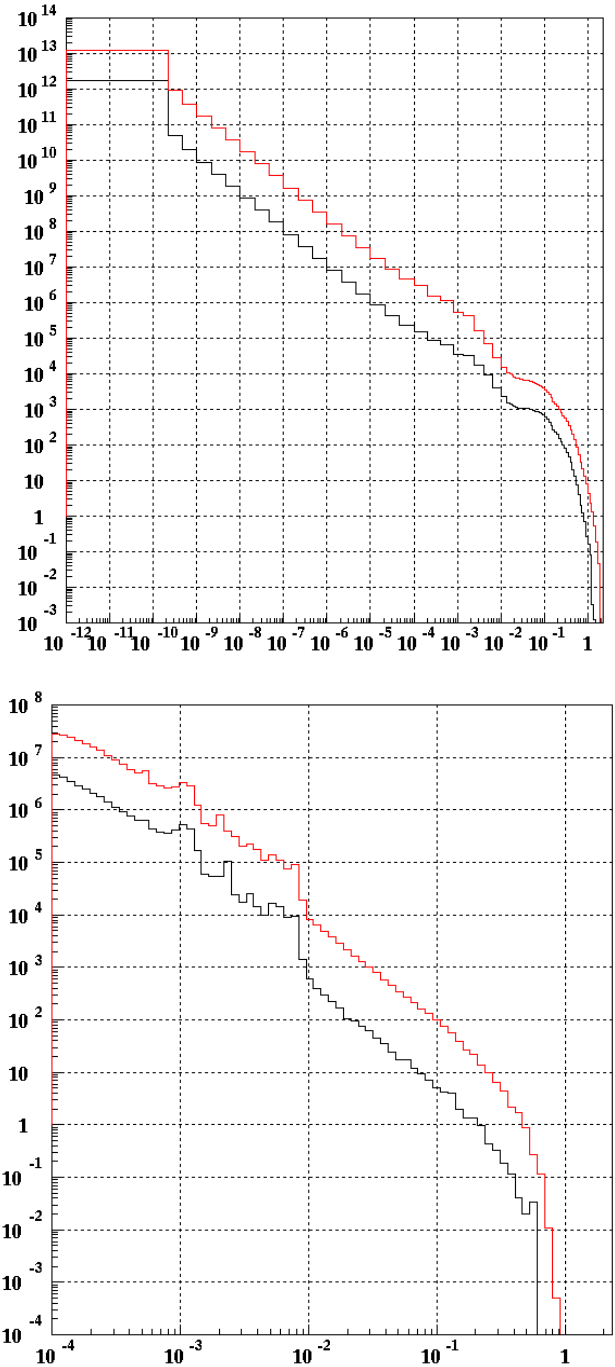

**Figure 13.** Energy spectra (cm$^{-2}$ GeV$^{-1}$ s$^{-1}$) vs energy (GeV) of neutrons (top) and gammas (bottom) averaged over a part of the third longitudinal section of the model (see Fig. 7) from z=550m to z=580m; the red and black lines represent the tunnel concrete and all the 10 transverse bins in soil, respectively.

## 6. Verification of predicted residual activation

A comparison with a relatively simple activation experiment performed at GSI, when the number of projectiles delivered to the target is well known, is presented in Ref. [3]. In this case, the MARS15 predictions agree with measured foil activation within 30% or better. This case, however, does not represent all the complexity of experiments one could encounter. Therefore, comparisons with activation measured in a more complicated experiment at Fermilab are presented below.

*6.1. Comparisons between measured and calculated residual activation and a normalization issue*
Residual activation of various aluminum, steel and copper foils was measured in a collimation region of Main Injector at Fermilab with 8-GeV primary proton beam [11-13]. In this experiment, foil activation was measured near the beam pipe (unshielded location) and on a collimator wall which represents a steel and marble shielding approximately 60 and 15 cm, respectively, in thickness (shielded location). The beam line and foil locations are shown in Figs. 14 and 15.

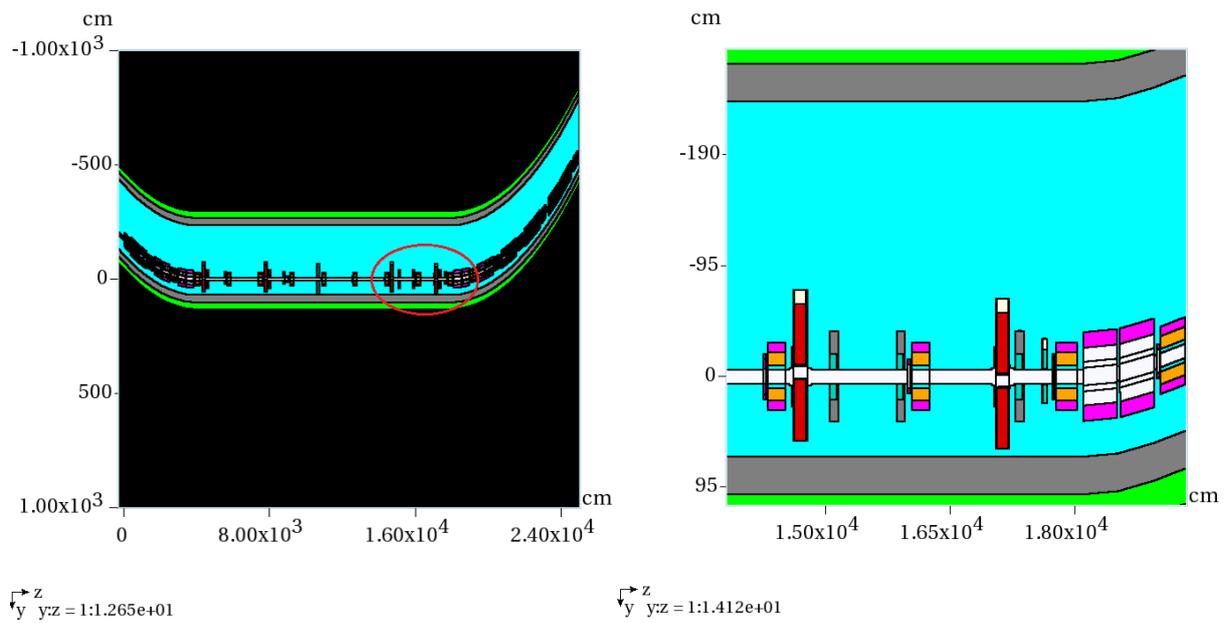

**Figure 14.** A plan view of the MARS15 model of the Main Injector collimation region (left) and the beam line fragment used for the measurements (right).

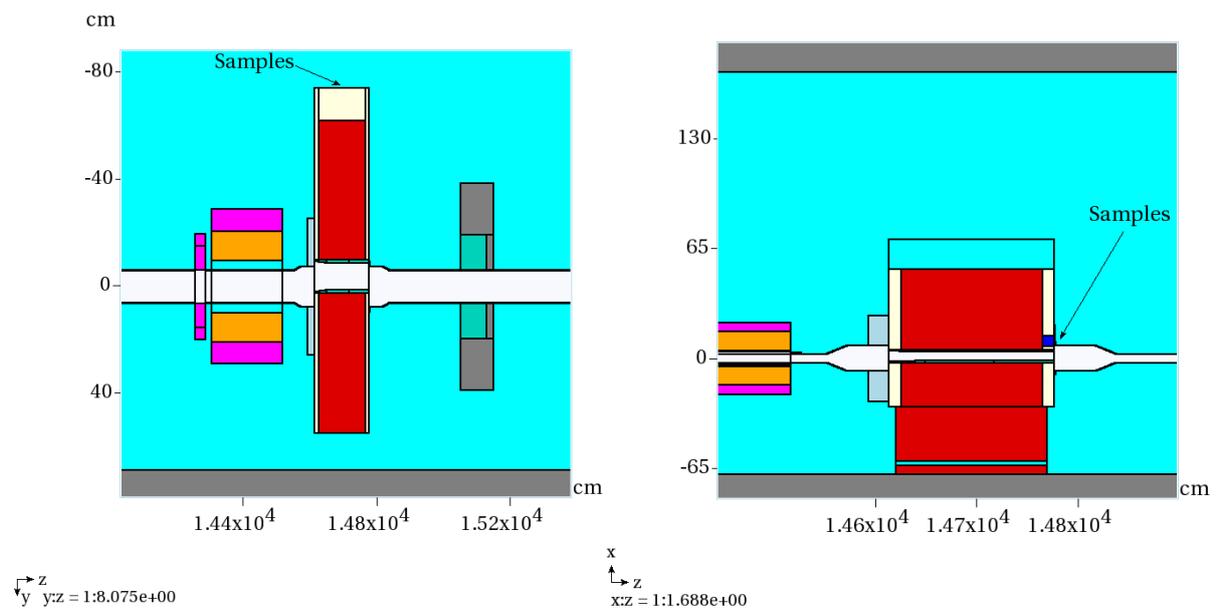

**Figure 15.** The beam line fragment with the activation foils: a plan view (left) and elevation view (right) that show the shielded and unshielded location, respectively.

Various irradiation and cooling times were studied. Three different methods were used in order to determine the number of protons lost on the collimator with the activation foils. The three methods provided three different numbers, with the ratios of both the highest number to the average and the average to the lowest number being approximately 1.4. So, the ratio of the highest number to the lowest one is approximately 2. In order to normalize the calculated foil activation, the average of the three estimated numbers is used. The detailed measured dependence of beam intensity vs time is used to normalize the calculated foil activation. Another source of potential uncertainty is that material composition of the steel samples is not well known, so that an average—over several available specifications— material composition is used. For aluminum and copper samples, corresponding natural mixtures are used. The comparisons between the measured and calculated foil activation are presented in Tables 7 thru 12.

*6.2. Unshielded location*

The steel and copper foils presented in this section were installed and removed together, so that the same irradiation and cooling down profile applies to both these foils. One can see that for most of the radionuclides in Tables 7-9 the agreement with measurement is within a factor of approximately 2, while for several radionuclides it is worse. Another conclusion from this comparison is that the calculations mostly **underestimate** the measured activation. It may be a consequence, in particular, of the above-mentioned systematic uncertainty in the number of protons that are lost on the collimator.

**Table 7.** Measured in Radionuclide Analysis Facility at Fermilab [11-13] and calculated specific activity (pCi/g) of various radionuclides generated in a steel foil irradiated at the unshielded location for about 93 hours.

| Produced radionuclide | Experiment | Calculation | C/E |
|---|---|---|---|
| $^{42}$K | 1510±520 | 1300 | 0.86 |
| $^{43}$K | 1240±170 | 955 | 0.77 |
| $^{48}$Cr | 863±112 | 1830 | 2.12 |
| $^{51}$Cr | 10700±2000 | 4860 | 0.45 |
| $^{52}$Fe | 904±159 | 645 | 0.71 |
| $^{52}$Mn | 12600±1100 | 11786 | 0.94 |
| $^{54}$Mn | 2200±380 | 808 | 0.37 |
| $^{24}$Na | 944±155 | 176 | 0.19 |
| $^{44}$Sc | 7680±1200 | 4230 | 0.55 |
| $^{46}$Sc | 463±103 | 218 | 0.47 |
| $^{47}$Sc | 3080±480 | 1635 | 0.53 |
| $^{48}$Sc | 702±129 | 860 | 1.22 |
| $^{48}$V | 4430±510 | 2776 | 0.63 |

**Table 8.** Measured in Radionuclide Analysis Facility at Fermilab [11-13] and calculated specific activity (pCi/g) of various radionuclides generated in an aluminium foil irradiated at the unshielded location for about 93 hours.

| Produced radionuclide | Experiment | Calculation | C/E |
|---|---|---|---|
| $^{7}$Be | 712±115 | 170.4 | 0.24 |
| $^{22}$Na | 355±55 | 115.0 | 0.32 |

**Table 9.** Measured in Radionuclide Analysis Facility at Fermilab [11-13] and calculated specific activity (pCi/g) of various radionuclides generated in a copper foil irradiated at the unshielded location for about 93 hours.

| Produced radionuclide | Experiment | Calculation | C/E |
|---|---|---|---|
| $^{55}$Co | 1510±290 | 2014 | 1.33 |
| $^{56}$Co | 642±145 | 252 | 0.39 |
| $^{57}$Co | 537±197 | 282 | 0.53 |
| $^{58}$Co | 3310±540 | 1532 | 0.46 |
| $^{61}$Cu | 47800±5000 | 25690 | 0.54 |
| $^{43}$K | 800±204 | 152 | 0.19 |
| $^{52}$Mn | 2490±240 | 1614 | 0.65 |
| $^{56}$Mn | 5400±720 | 2129 | 0.39 |
| $^{57}$Ni | 1270±250 | 1113 | 0.88 |
| $^{44}$Sc | 1470±280 | 1126 | 0.77 |
| $^{47}$Sc | 501±201 | 404 | 0.81 |
| $^{48}$Sc | 413±82 | 377 | 0.91 |
| $^{48}$V | 519±107 | 506 | 0.97 |

*6.3. Shielded location*

The steel, aluminium and copper foils presented in this section were installed and removed together, so that the same irradiation and cooling down profile applies to both these foils. One can see that for most of the radionuclides in Tables 10-12 the agreement with measurement is within a factor of approximately from 2 to 3, while for $^{56}$Co it is about 4. Another conclusion from this comparison is that the calculations mostly **overestimate** the measured activation. It may be a consequence, in particular, of the above-mentioned uncertainty in the number of protons that are lost on the collimator.

**Table 10.** Measured in Radionuclide Analysis Facility at Fermilab [11-13] and calculated specific activity (pCi/g) of various radionuclides generated in a steel foil irradiated at the shielded location for about 45 days.

| Produced radionuclide | Experiment | Calculation | C/E |
|---|---|---|---|
| $^{56}$Co | 22.5±4.3 | 99.6 | 4.43 |
| $^{51}$Cr | 1870±290 | 3504 | 1.87 |
| $^{52}$Mn | 439±40 | 1548 | 3.53 |
| $^{54}$Mn | 639±97 | 1520 | 2.38 |
| $^{46}$Sc | 58.3±7.7 | 145.6 | 2.50 |
| $^{47}$Sc | 44.9±11.6 | 100.1 | 2.22 |
| $^{48}$V | 358±39 | 880 | 2.46 |

**Table 11.** Measured in Radionuclide Analysis Facility at Fermilab [11-13] and calculated specific activity (pCi/g) of various radionuclides generated in an aluminium foil irradiated at the shielded location for about 45 days.

| Produced radionuclide | Experiment | Calculation | C/E |
|---|---|---|---|
| $^{22}$Na | 45.8±11.8 | 41.7 | 0.91 |
| $^{24}$Na | 1270±210 | 1102 | 0.87 |

**Table 12.** Measured in Radionuclide Analysis Facility at Fermilab [11-13] and calculated specific activity (pCi/g) of various radionuclides generated in a copper foil irradiated at the shielded location for about 45 days.

| Produced radionuclide | Experiment | Calculation | C/E |
|---|---|---|---|
| $^{56}$Co | 78.7±8.2 | 307 | 3.90 |
| $^{57}$Co | 138±18 | 324 | 2.35 |
| $^{58}$Co | 707±107 | 1844 | 2.61 |
| $^{60}$Co | 22.6±3.5 | 50.4 | 2.23 |
| $^{51}$Cr | 131±51 | 256 | 1.95 |
| $^{59}$Fe | 70.0±9.4 | 50.1 | 0.72 |
| $^{52}$Mn | 43.9±5.2 | 128 | 2.92 |
| $^{54}$Mn | 49.3±8.4 | 85.4 | 1.73 |
| $^{48}$V | 39.1±5.0 | 36.6 | 0.94 |

## 7. Conclusions

Spatial distributions of radionuclides produced in the soil outside the ESS accelerator tunnel wall were calculated for a realistic operational scenario. The distributions are for subsequent analysis of radionuclide migration by a dedicated expert group. A comparison between predicted and measured foil activation, performed for a similar radiation environment, for most of the studied radionuclides revealed an agreement between calculated and measured residual foil activation within a factor of from one to three.

**Acknowledgements**
This document was prepared using the resources of the Fermi National Accelerator Laboratory (Fermilab), a U.S. Department of Energy, Office of Science, HEP User Facility. Fermilab is managed by Fermi Research Alliance, LLC (FRA) acting under contract No. DE-AC02-07CH11359.
    This research used a computing award allocation from the ASCR Leadership Computing Challenge at the Argonne Leadership Computing Facility, which is a DOE Office of Science User Facility supported under Contract DE-AC02-06CH11357.
    This work was supported by a CRADA agreement NO FRA-2016-0058 between Fermi Research Alliance, LLC and European Spallation Source ERIC.
    Authors are grateful to B. C. Brown of Fermilab for helpful discussions.

**References**
[1]    https://europeanspallationsource.se/accelerator
[2]    Mokhov N V 1995 The MARS Code System User's Guide *Fermilab-FN-628*
[3]    Mokhov N V *et al* 2014 *Prog. Nucl. Sci. Technol.* **4** 496
[4]    https://mars.fnal.gov
[5]    Aarnio P 1998 Decay and Transmutation of Nuclides *CERN CMS-NOTE-1998/086*

[6]  Mokhov N, Rakhno I, Tropin I, Eidelman Yu and Tchelidze L 2016 *ESS accelerator prompt radiation shielding assessment ESS-0052477, rev. 2*
[7]  Mokhov N V, Eidelman Yu I, Rakhno I L, Tchelidze L, Tropin I S 2016 MARS15 simulation of radiation environment at the ESS linac *Fermilab-Conf-16-543-APC*
[8]  Krivosheev O E and Mokhov N V 2000 Tolerable Beam Loss at High-Intensity Proton Machines *Fermilab-Conf-00/192*
[9]  ESS hands on maintenance conditions for ESS accelerator *ESS-0008351, rev. 2*
[10] Thomas R H and Stevenson G R 1988 Radiological Safety Aspects of the Operation of Proton Accelerators *IAEA TERS 283*
[11] Brown B C 2011 Analysis procedures for Al activation studies *Fermilab Beams-doc 3980*
[12] Brown B C 2015 Calibration of BLM response to proton loss *Fermilab Beams-doc 4867*
[13] Brown B C 2017 Activation of steel and copper samples in the Main Injector collimation region *Fermilab Beams-doc 4046 ver. 2.27*